\documentclass[a4paper,11pt]{article}
\usepackage{jheppub} % for details on the use of the package, please see the JINST-author-manual

\arxivnumber{2501.06687} % if you have one

\usepackage{amssymb,amsthm,amsmath}
\usepackage{textcomp}
\usepackage{color}      % color is used in text
\usepackage{slashed}    % Feynman slash
\usepackage{verbatim}
\usepackage{epsfig}
\usepackage[normalem]{ulem} 
\usepackage{rotating}   % to rotate tables
\usepackage{multirow}   % multicolumn and multirow

\usepackage{bm}
\renewcommand{\figurename}{Fig.}

\renewcommand{\tablename}{Table}

\newcommand{\hatp}{\hat{\bm p}}
\newcommand{\hatk}{\hat{\bm k}}

\newcommand{\hatl}{\hat{\bm l}}

\begin{document}
	\raggedbottom
	
	\title{
		Precise measurement of CP violating $\tau$ EDM through $e^+ e^- \to \gamma^*, \psi(2s) \to \tau^+ \tau^-$ 
	}

	\author[a,b]{Xiao-Gang He
	}\emailAdd{hexg@sjtu.edu.cn}
	\affiliation[a]{State Key Laboratory of Dark Matter Physics, Tsung-Dao Lee Institute and School of Physics and Astronomy, Shanghai Jiao Tong University, Shanghai 201210, China} 
	\affiliation[b]{Key Laboratory for Particle Astrophysics and Cosmology (MOE) \& Shanghai Key Laboratory for Particle Physics and Cosmology, Tsung-Dao Lee Institute and School of Physcis and Astronomy, Shanghai Jiao Tong University, Shanghai 201210, China}
	
	\author[a,b]{   Chia-Wei Liu  
	}\emailAdd{chiaweiliu@sjtu.edu.cn}

	\author[c,d,e]{Jian-Ping Ma}\emailAdd{majp@itp.ac.cn}
	\affiliation[d]{School of Physics, Henan Normal University, Xinxiang 453007,Henan, China}	
	\affiliation[e]{Institute of Theoretical Physics, P.O. Box 2735, Chinese Academy of Sciences, Beijing 100190, China}
	\affiliation[f]{School of Physics and Center for High-Energy Physics, Peking University, Beijing 100871, China}

	\author[a,b]{    
		Chang Yang 
	}
	\emailAdd{15201868391@sjtu.edu.cn
	}

	\author[a,b]{   
		Zi-Yue Zou
	}\emailAdd{ziy\_zou@sjtu.edu.cn}

	\date{\today}

\abstract{
A nonzero electric dipole moment of a tauon, \(d_\tau\), signals CP violation and provides an important probe for new physics.  We study methods to measure \(d_\tau\) at  low energy $e^+ e^-$ colliders  through the processes $e^+e^- \to \gamma^*, \psi(2S) \to \tau^+\tau^-$ with $\tau^\pm$ decays into a charged hadron and a tau neutrino.  We point out that, with   measuring energies of the charged hadron, Im$(d_\tau)$ can be measured. On the other hand,  selecting events of $\tau$ decays after traveling more than the detector resolution distance, Re$(d_\tau)$ can also be determined. We find that the precision at Super Tau-Charm Facility (STCF) running at  the center energy of $m_{\psi (2S)}$ for 10 year data accumulation, the precision of Im$(d_\tau)$ and Re$(d_\tau)$ are found to be  1.8  and 11 in unit of $ 10^{-18}~e\,\text{cm}$, respectively. The sensitivity for \(d_\tau\) measurement precision at the STCF can be reached its optimum at a central energy of \(6.3~\text{GeV}\), achieving a precision of $0.7$   for Im$(d_\tau)$ and $2.8$  for Re$(d_\tau)$ in unit of $ 10^{-18}~e\,\text{cm}$. 
	}

	\maketitle
	
	\section{Introduction} 
	The electric dipole moment (EDM) of a fundamental fermion   violates CP symmetry. In the Standard Model (SM),   EDMs are
	generated only through higher-order loop processes and 
	therefore are predicted to be extremely small~\cite{hmp,Bernreuther:1990jx,Chupp:2017rkp,Yamaguchi:2020eub}. Experimental searches for EDMs have so far yielded null results~\cite{Muong-2:2008ebm,ACME:2018yjb,ParticleDataGroup:2024cfk,OPAL:1996dwj}, with some particles having poorly constrained limits. In particular, for the tauon  EDM, \(d_\tau\), the most stringent constraint  obtained at Belle is found to be~\cite{Belle:2021ybo}
	\begin{eqnarray}\label{bel}
		\text{Re}(d_\tau)
		&=& (-6.2 \pm 6.3) \times 10 ^{-18} e\text{cm}\,, \nonumber\\
		\text{Im}(d_\tau)
		&=& (-4.0 \pm 3.2) \times 10 ^{-18} e\text{cm} \,. 
	\end{eqnarray}
	New physics beyond the SM could potentially generate a   significantly larger  \(d_\tau\) that might be detectable in the near future~\cite{Bernreuther:2021elu}.  
	Due to the short lifetime of tauons, measuring their EDM in the light-like region is challenging~\cite{Huang:1996jr,Bernreuther:1996dr}.
	In this study, we explore the sensitivity reach for \(d_\tau\) through the  process \(e^+e^- \to \gamma^*, \psi(2S) \to \tau^+\tau^-\) at the Super Tau Charm Facility (STCF)~\cite{Achasov:2023gey} with the intermediate states $\gamma^*$ and $\psi(2S)$, aiming to better constrain physics beyond the SM. 
	At STCF, the energy of the $\tau$ pair produced is not high, and the distance traveled may be short, making the reconstruction of the $\tau$ momentum challenging and limiting the information needed for $d_\tau$ extraction. We propose strategies to overcome these difficulties.
	Our findings indicate that the sensitivity for \(d_\tau\) at the STCF could achieve tighter limits than the current best experimental constraints.
	
	The cross-sectoion of \( e^+ e^- \to \gamma ^* \to  \tau^+ \tau^- \), at the leading order,  is given by  
	\begin{equation}
		\sigma = 
		\frac{4\pi \alpha _{\text{em}} ^2 }{3s} \sqrt{
			1 - \frac{4 m^2_\tau  }{s}
		}\left(
		1+ \frac{2 m^2_\tau  }{s}
		\right)\,,
	\end{equation}  
	Here, \(\sqrt{s}\) is the center-of-mass energy of the system, and \(\alpha_{\text{em}}  = e ^2 / ( 4 \pi ) \) is the electromagnetic fine structure constant.  
	At \(\sqrt{s} = m_{\psi(2S)}\),  
	the photon propagator receives an enhancement from $\gamma ^* \to \psi (2S) \to \gamma ^* :  $
	\begin{equation}
		\frac{1}{s}   \to \frac{1}{s}  + \frac{4\pi\alpha_{\text{em}}  Q_c^2  g^2_{\psi(2S)} }{    i \sqrt{s}^5  \Gamma_\psi  }  ,
	\end{equation}
	where  $Q_c=2 /3 $ and $\Gamma_\psi $ is the  width of $\psi (2S)$.  The decay constant $g_{\psi(2S)}$ is given by  
	$
	\langle \psi (2S) | \overline{c} \gamma_\mu c | 0 \rangle = g_{\psi(2S)} \epsilon_\mu$  with  \(\epsilon_\mu\) being the polarization vector of \(\psi(2S) \). It can be determined from $\psi(2S) \to \tau^+\tau^-$ branching ratio of $(3.1\pm 0.4 )\times 10^{-3}$. 
	Numerically, at \(\sqrt{s} = m_{\psi(2S)}\), the cross section \(\sigma\) is enhanced from 2.5~nb to 4.5~nb.  
	At STCF, the luminosity is expected to reach \(1~\text{ab}^{-1}\) per year. Over ten years of data collection, the total number of \(\psi(2S)\) events is anticipated to be about \(2 \times 10^{10}\)~\cite{Achasov:2023gey}.  
	
	Information about the $\tau$ EDM $d_\tau$ originates from the interaction Lagrangian of
	\begin{eqnarray}
		L_{\text{int}} = -i \frac{1}{2} d_\tau(q^2) \bar{\tau} \sigma_{\mu\nu} \gamma_5 \tau F^{\mu\nu},
	\end{eqnarray}
	where $F^{\mu\nu}$ is the photon field strength and $q$ is the momentum of $\gamma^*$.  At \(q^2 = 0\), \(d_\tau(0)\) represents the usual tau EDM and must be real.  
	For the process \(e^+e^- \to \gamma^* \to \tau^+\tau^-\), \(d_\tau\) is evaluated at the energy scale \(q^2 = s\), where \(d_\tau\) also develops an imaginary part, \(\mathrm{Im}(d_\tau)\).	
	
	To measure the EDM and test CP symmetry in the relevant process, one needs to extract information about the \(\tau^\mp\) spins~\cite{BLMN}.  
	The spin effects are reflected in the hadrons during the sequential decays:
	\begin{eqnarray}
		\frac{
			d \Gamma (\tau^-  \to h^-    \nu_\tau ) 
		}{
			d \cos \theta_- 
		}
		&=& \frac{1}{2} (1 +
		\alpha_{h  } 
		\cos \theta_-)  \,,
		\nonumber\\
		\frac{
			d \Gamma (\tau^+  \to h^{ +}   \overline{\nu}_\tau )
		}{
			d \cos \theta_+ 
		}
		&=&  \frac{1}{2} (1 - 
		\overline{\alpha}_{h} 
		\cos \theta_+) \,.
	\end{eqnarray}  
	In this work we consider the cases of $h^{ \pm }  = \pi^\pm$ or $\rho^{\pm}$.  
	Taking CP  to be  conserved in the   cascade decays would lead to $
	\overline{\alpha}_h  = \alpha _ h . 
	$  
	Here, \(\theta_-(\theta_+)\) represents the angle  between the \(\tau^-(\tau^+)\) spins and the 3-momenta ${\bm l}_-({\bm l}_+)$ of the outgoing secondary hadrons \(h^-(h^{+})\) in the rest frame of \(\tau^- (\tau^+) \).
	In the SM, neutrinos are left-handed, and  the helicities in \(\tau^- \to h^-\nu_\tau\) are fixed by the $V-A$ structure. The helicity-related parameters are determined as \((\alpha_\pi, \alpha_\rho) = (1, 0.45)\)~\cite{Bernreuther:2021elu}.
To improve the statistics, it would be useful to include more subsequent decay channels~\cite{Bernreuther:2021uqm}. To demonstrate our novel idea for the selection rule, we will focus exclusively on the two-body   decays of $\tau^{\mp}$.
	%In this work, we extract the impact of \(d_\tau\) on the angular distributions of the final states.

	\section{Measurements of  Im$(d_\tau)$ and Re$(d_\tau)$}
	
	By measuring ${\bm l}_\pm$  and the three-momentum ${\bm k}$ of \(\tau^-\),  the \(\tau\)-EDM can be extracted.  
	The imaginary and real parts of \(d_\tau\) can be separately determined using the combinations of observable quantities discussed below~\cite{Du:2024jfc,Du:2024jfc2}.

	For the imaginary part, we have
\begin{equation}
\text{Im} (d_\tau)=  	\frac{	-e (  3s +  6 m_\tau^2 )}{4 m_\tau \sqrt{s}  \sqrt{s -4 m_\tau^2 } }
\left(\frac{\langle \hatl_- \cdot \hatk \rangle }{\alpha_h }+ \frac{\langle\hatl_{+}    \cdot \hatk \rangle }{\overline{\alpha} _{h'}  }\right), 
\end{equation}
	where $\hatl_\pm$ and $\hatk$ are unit vectors for the directions of the momenta ${\bm l}_\pm$ and ${\bm k}$, respectively. 
	There are two different methods to extract the real part of the EDM from the distributions:  
	\begin{equation}\label{1re}
		\text{Re}(d_\tau)^a
		= e
		\frac{9}{4}
		\frac{
			s + 2 m_\tau^2 
		}{
			\alpha_h \overline\alpha_{h'} m_\tau  
			\sqrt{
				s^2 - 4 s m_\tau^2 
			} 
		}
		\langle
		(\hatl_- \times 
		\hatl_+) \cdot \hatk 
		\rangle \,, 
	\end{equation}
	and 
	\begin{equation}\label{2re}
		\text{Re}(d_\tau)^b
		=  e
		\frac{45}{2}
		\frac{
			(s + 2 m_\tau^2) \langle
			(\hatp \cdot \hatk) 
			(\hatl_- \times 
			\hatl_+) \cdot \hatp
			\rangle 
		}{
			\alpha_h \overline\alpha_{h'} 
		\sqrt{s} (\sqrt{s} - 2 m_\tau) 
			\sqrt{
				s - 4 m_\tau^2 
			} 
		}  \,.
	\end{equation}
	The superscripts \(a\) and \(b\) denote the first a) and second b) methods, respectively. $\hatp$ is the unit vector 
	for the moving direction of the initial  $e^-$. 
	The brackets $\langle {\cal O} \rangle$ denote the average value of ${\cal O}$ over the entire angular distribution.

	In the above, $\tau^-$ decays to the hadron $h^-$, and $\tau^+$ decays to the hadron $h^{\prime +}$. It is noted  that $h^-$ and $h^{\prime +}$ are not necessarily the same type of  hadrons. To achieve the best precision, we have to  consider different permutations of hadrons from $\tau^+$ and $\tau^-$ decays and take the average of the measurements. 
	
 \subsection{Measurement of Im$(d_\tau)$} 
	
	For Im$(d_\tau)$ measurement, one needs to know $\langle \hatl_{\mp}   \cdot \hatk \rangle$. The inner products are related to $h^-$ and $h^{\prime +}$ energies $E_\mp$, respectively, in the lab frame as 
	\begin{equation}\label{he}
\boldsymbol{\hat{l}_{\pm} \cdot \hat{k}} = \pm \frac{4E_{\pm} m_{\tau}^2 / \sqrt{s} - m_h^2 - m_{\tau}^2}{(m_{\tau}^2 - m_h^2) \sqrt{1 - 4m_{\tau}^2 / s}}
	\end{equation}
	%Here $E_\mp$ are the energies of $h^-$ and $h^{\prime +}$ in the lab frame, respectively.  
	It is interesting to note that the measurements of \(\text{Im}(d_\tau)\)   
	require only the detection of  energies of hadrons but not   full three-momenta of $\tau$'s and hadrons. Once the $e^+e^-$ center of mass frame energy $\sqrt{s}$ is known, one only needs to measure $E^{\pm}$ to obtain  $\langle \hatl_{\pm}   \cdot \hatk \rangle$. 
	
	The sensitivity $\delta_{\text{Im}}$ for the measurement of Im$(d_\tau)$ can be estimated as the following. From an observable ${\cal O}$, in general the standard deviation $\delta _ {\cal O}$ is given by $ \sqrt{( \langle {\cal O} ^2  \rangle -  \langle {\cal O}   \rangle ^2) / N }$  with  $N$ the   number of events. We have  the standard deviation 
	of Im$(d_\tau)$ as 
\begin{equation}\label{10}
\delta _{\text{Im}} = 	\frac{e (			s +2  m_\tau^2)}{4 m_\tau  \sqrt{s}   \sqrt{s -4 m_\tau^2 } 	} \sqrt{\frac{3 }{  \displaystyle \sum _h^{\pi , \rho }     \alpha _ h ^2 	N_{\text{Im}}^{h}   }+\frac{3}{  \displaystyle \sum _h^{\pi , \rho }     	  \overline{\alpha}^2 _h 		\overline{N} _{\text{Im}}^{h} }}\,. 
\end{equation} 
	The   event number is given by 
	\begin{eqnarray}
		N_{\text{Im}}^{h} 
		&=& \epsilon  
		L 
		\sigma  {\cal B} ( \tau ^- \to h^ - \nu _\tau )  \,,\nonumber\\ 
		\overline{N} _{\text{Im}}^{h} 
		&=&\epsilon  
		L 
		\sigma  {\cal B} ( \tau ^+ \to h^ + \nu _{\overline{\tau} }  )  \,,
	\end{eqnarray}
	where \(\epsilon  \) is the detection efficiency and \(L\) the luminosity of 
	\(e^- e^+\) collisions. 
	The factor \(\sqrt{s - 4 m_\tau^2}\) in the denominator is crucial. It indicates that \(\text{Im}(d_\tau)\) is difficult to measure at \(\sqrt{s} \sim 2 m_\tau\) and degrades the precision at \(\sqrt{s} = m_{\psi(2S)}\). In practice,   \(\alpha_h^2\) can be interpreted as the efficiency of reconstructing spins from momentum.  
	Although  
	\({\cal B} (\tau^- \to \pi ^- \nu_\tau) = (10.82 \pm 0.05)\%\) is smaller than  
	\({\cal B} (\tau^- \to \rho ^- \nu_\tau) \approx 25\%\)~\cite{ParticleDataGroup:2024cfk}, the channel with \(h = \pi\) contributes twice as much statistically significant data due to \(\alpha_\rho^2 / \alpha_\pi^2 \approx 0.2\) as evidenced in Eq.~\eqref{10}.
	In the following, we neglect CP violation in \(\tau^- \to h^- \nu_\tau\) and take \(\alpha_h = \overline{\alpha}_h\).
	\\
	
 \subsection{Measurement of Re$(d_\tau)$} 
	
	The measurement of Re$(d_\tau)$ needs however, in both ways a) and b) mentioned earlier, the full reconstruction of   $\hatk$ and $\hatl_\mp $. The momentum directions of the secondary hadrons  can be measured in the experiment, but the measurement of $\hatk$ is much more involved.
	
	With measured $\hatl_\pm$ one can partly 
	reconstruct $\hatk$.  This leads to the formula~\cite{Belle:2021ybo,OPAL:1996dwj}:
	\begin{eqnarray}\label{kha}
		\hatk &=&  
		u \hatl_+  + v  \hatl_- + \text{sgn} \left(
		(\hatl_- \times \hatl_+) \cdot \hatk
		\right) w  \hatl_- \times \hatl_+ \,.
	\end{eqnarray}
	In the above,  \(u\) and \(v\) are functions of \(\hatl_+ \cdot \hatl_-\) and can be obtained by matching to Eq.~\eqref{he}, while \(w\) is determined by  \(|\hatk | = 1\) and positive,  only  the sign of   $(\hatl_- \times \hatl_+) \cdot \hatk)$ in the last term can not be fixed. This is  the two-fold ambiguity in reconstruction of \(\hatk\) from \(\hatl_\pm\)  due to the undetected neutrinos.  
	In some literature,  the sign is treated as a random number taking either \(\pm 1\)~\cite{Sun:2024vcd,Belle:2021ybo}. This approach may suffice for measuring \(\text{Im}(d_\tau)\), as \((\hatl_- \times \hatl_+) \cdot \hatl_\mp = 0\). Treating  the sign  as a random number  of $+1$ or $-1$   leads to  $\langle (\hatl_- \times \hatl_+) \cdot \hatk \rangle = 0 $, because the expectation value of the random number vanishes. Hence, it is impossible to perform a  measurement of  \(\text{Re}(d_\tau)^a\)  as long as \(\hatk\) itself is not measured~\footnote{ Similar work on \(\tau\) EDM measurement at the STCF has been carried out recently in Ref.~\cite{Sun:2024vcd}. There, instead of identifying observables to isolate the \(d_\tau\) effects, they fit the full angular distribution to extract \(d_\tau\). For the measurement of \(\text{Im}(d_\tau)\), we obtain similar results, but their determination of \(\text{Re}(d_\tau)\) suffers from the ambiguity problem mentioned here.
		In addition, the case of \(h = \pi\), with higher statistical significance, was not considered in Ref.~\cite{Sun:2024vcd}.
	}.
	Also the ambiguity in the sign function also modifies \(\text{Re}(d_\tau)^b\). Therefore, for Re$(d_\tau)$, measurements of \(\hatl_\pm\) alone are not sufficient.
	To address this shortcoming, we propose fully reconstructing \(\hatk\) in future CP tests by selecting $\tau$ decay events   traveling  more than the detector spacial resolution length. This procedure will sacrifice the statistic, but as will be seen later, at the STCF, good sensitivities can still be achieved.
	
	In symmetric colliders, such as the BESIII and  STCF, the momenta of charged particles can be measured if their flight distance surpasses the  resolution length $D$. We note that  it suffices for   measurements to determine 
	sgn$((\hatl _- \times \hatl _+) \cdot \hatk) $ for reconstructing $\hatk$. 
	The proportion of $\hatk$  being measured is then given by 
	\begin{equation}
		P_\tau 
		= 1 - \left(  
		\int_0^{
			D  / D_0  } 
		\exp \left( - x 
		\right) 	dx 
		\right)^2 ,
	\end{equation}
	where \(D_0 = \tau_{\tau }\sqrt{s/(4m^2_\tau) - 1}  \), representing the mean decay length of   \(\tau\), and $\tau_{\tau }$ the lifetime of $\tau$. The integral represents the probability of \(\tau^-\) decaying before its flight distance reaches \(D\) in the lab frame, and the square arises because it is sufficient to probe the momentum of either \(\tau^-\) or \(\tau^+\).
We note that it is also possible to reconstruct $\hat {\bm k}$ by measuring the impact parameter  between $h^+$ and $h^-$~\cite{Kuhn:1993ra}. Nevertheless, the precision of the impact parameter is also limited by the experimental spatial resolution. A portion of events with small impact parameters should be omitted, and the impact of this cut would complicate the measurements of the EDM.

	Note that for identifying the $\tau$ momentum direction, a smaller $D/D_0$ is preferable, as it leads to a larger reconstruction rate of $P_\tau$. Two approaches to achieve this are: (1.) increasing the energy of the $\tau$ in the laboratory frame, i.e., increasing the value of $s$; and (2.) enhancing the detector's spatial resolution to be finer than $D_0$.

	The standard deviations of 
	Re$(d_\tau)^{a,b}$   in order are 
	\begin{eqnarray}\label{1reR}
		\delta_{
			\text{Re} }(D)
		^a 	= 
		\frac{3 e}{4}
		\frac{
			s + 2 m_\tau^2 
		}{
			m_\tau  
			\sqrt{
				s^2 - 4 s m_\tau^2 
			} 
		}
		\sqrt{ 
			\frac{2}{N^{\text{eff}}_{\text{Re}}} 
		}
		\,,
	\end{eqnarray}
	and 
	\begin{eqnarray}\label{2reR}
		\delta_{
			\text{Re} }(D) 
		^b 	= \frac{3 e}{2}
		\frac{
			\sqrt{s^2 + 3 s m_\tau^2 + 2 m_\tau^4} 
		}{
\sqrt{s} (\sqrt{s} - 2 m_\tau) 
			\sqrt{
				s - 4 m_\tau^2 
			}
		}
		\sqrt{ 
			\frac{20}{N^{\text{eff}}_{\text{Re}}} 
		} \,, 
	\end{eqnarray}
	where the effective number of events is   given by 
	\begin{equation}\small 
		N_{\text{Re}}^{\text{eff}} 
		= P_\tau 
		\epsilon
		L  \sigma 
		\left(
		\sum _ {h}^{  \pi , \rho } 
		\alpha_ h ^2 	{\cal B}(\tau^- \to h ^- \nu_\tau) 
		\right)^2  .
	\end{equation}
	We have written out explicitly that $\delta _{\text{Re}}$ depends on the spatial resolution $D$.  
	%The factor \(\sqrt{s} - 2m_\tau\) in the denominators plays a crucial role and demonstrates that 
	%the EDM is difficult to measure with small released energies.  
	Comparing \(\delta_{\text{Re}}(D)^a\) and \(\delta_{\text{Re}}(D)^b\), it is evident that the first method, which requires the full reconstruction of \(\hatk\), achieves significantly better precision.
	In the following, we consider both methods for measuring \(\text{Re}(d_\tau)^{a,b}\) to reduce uncertainties. This results in a combined weighted error given by  
	\begin{equation}
		\frac{1}{\delta_{\text{Re}}(D)^2} = \frac{
			1
		}{
			(\delta_{\text{Re}}(D)^a)^2
		} 
		+ \frac{
			1
		}{
			(\delta_{\text{Re}}(D)^b)^2
		}\,.
	\end{equation}
	
	\section{Numerical results and discussions}
	
	\begin{figure}[t]
		\begin{center}
			\includegraphics[width=0.45 \linewidth]{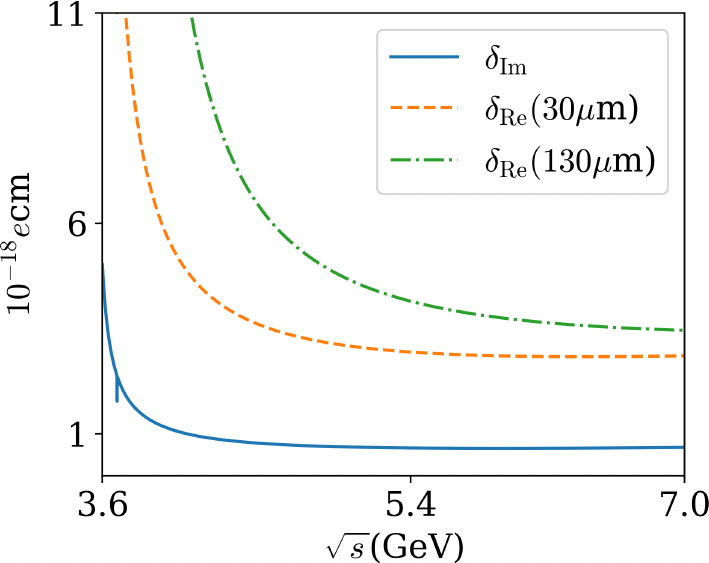}
			\caption{The  expected  precision of  $d_\tau$ 
				with \(L\epsilon = 0.63\,\text{ab}^{-1}\). 
			The values of $D$ for $\delta_{\text{Re} }$ are indicated in parentheses.  }
			\label{fig:two}
		\end{center}
	\end{figure}

	For \(\text{Im}(d_\tau)\), it is sufficient to measure the secondary hadron energies from $\tau$ decays to perform the measurement.  
	At BESIII, the number of events for the process \(\psi(2S) \to \tau^- \tau^+\) is approximately \(9 \times 10^6\)~\cite{BESIII:2024lks}, and we take \(\epsilon = 6.3\%\) for efficiencies\footnote{
	\(\epsilon = 6.3\%\) is the reported signal efficiency 
	 at STCF 
	for \( h = \rho \) quoted in Ref.~\cite{Sun:2024vcd}, where various efficiencies, including tracking, particle identification for photons, and reconstruction algorithm efficiencies, have been considered, among others. For \( h = \pi \), the efficiency should be higher due to fewer reconstructed final states, but we conservatively assume \( \epsilon = 6.3\% \) as well.
}.  It results in 
$\delta_{\text{Im}}  = 1.0 \times 10 ^{-16}e\text{cm} $ at BESIII. The sensitivity is not compatible with the current best value~\cite{Belle:2021ybo}.
	
	STCF is planned to operate with an energy range of 2.0--7.0~GeV, delivering an annual integrated luminosity of 1~ab\(^{-1}\)~\cite{Achasov:2023gey}. We assume a data sample collected for 10 years  
	and use \(\epsilon = 6.3 \% \)~\cite{Sun:2024vcd}  for our numerical estimates. 
	The dependencies of $\delta_{\text{Re} }$ and $\delta_{\text{Im} }$ on \(\sqrt{s}\) are depicted in Fig.~\ref{fig:two}.	For the measurement of $\mathrm{Re}(d_\tau)$, a spatial resolution of $D = 130\,\mu $m has already been achieved at BESIII and is expected to be achieved at the  STCF. If a silicon pixel detector is implemented, the spatial resolution $D$ can be improved to $30\,\mu  $m, which is also used in our estimates.
	A bump is observed around \(3.7~\text{GeV}\) due to the \(\psi(2S)\) resonance.  
	We see that the best place to probe \(\text{Im}(d_\tau)\) is around \(\sqrt{s} = 6.3~\text{GeV}\), where the precision can reach 
 $0.7 \times 10^{-18}~e \text{cm},$ four times   better than the current value.

	\begin{figure}[tbh] 
		\begin{center}
			\includegraphics[width=0.45 \linewidth]{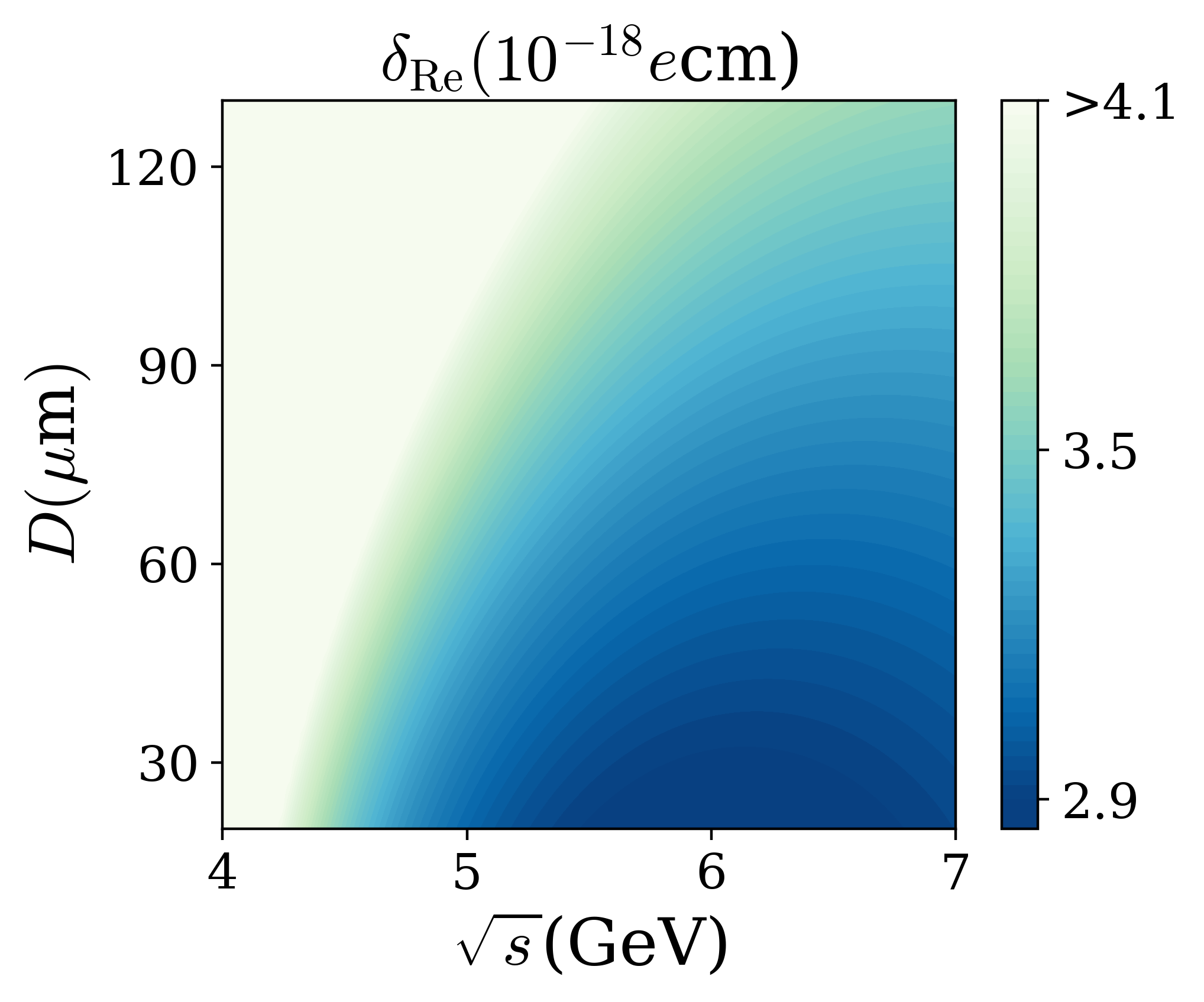}
			\caption{The precision of \(\text{Re}(d_\tau)\) may be achieved with \(L\epsilon = 0.63 \,\text{ab}^{-1}\). The color indicates the values of \(\delta_{\text{Re}}\), as shown in the color bar on the right.
			}
			\label{fig:three}
		\end{center}
	\end{figure}

	\begin{table}[tbh]
		\caption{The precision of \( d_\tau \) that may be achieved with \( L\epsilon = 0.63 \,\text{ab}^{-1} \) is given in units of \( 10^{-18} e\,\text{cm} \). The absolute value is defined as 
			$		\delta_{|d_\tau|}^2(D) = \delta_{\text{Re}}(D)^2 + \delta_{\text{Im}}(D)^2,
			$		where \( D \) is in units of \(\mu\text{m}\). The case \( D = 0 \) corresponds to situations where the \(\tau\)-lepton momentum can be reconstructed with 100\% accuracy which is shown only as a reference number.
		}
		\label{reedm}
\centering
		\begin{tabular}{|c|cccccc|}
			\hline
			\hline   
			$\sqrt{s}$ 
			&  $m_{\psi(2S)}$ &4.2 GeV  &4.9 GeV  & 5.6 GeV &6.3 GeV  &7 GeV 
			\\ 
			\hline
			$\delta_{\text{Im}}$ & 1.8 & 0.9 &{  0.7}&{ 0.7}&{  0.7}&{  0.7}\\
			$\delta_{\text{Re}}(  
			180)$&235&14.7&6.5&4.9&4.2&4.0\\
			$\delta_{\text{Re}}(  
			130)$&83&9.4&5.0&4.0&3.6&3.5\\
			$\delta_{\text{Re}}(  
			80)$&29&6.2&3.9&3.3&3.1&3.1\\
			$\delta_{\text{Re}}(  
			30)$&11&4.4&3.2&2.9&2.8&2.9\\
			$\delta_{\text{Re}}(  
			0)$&7.7&4.0&3.0&2.8&2.8&2.8\\
			$\delta_{|d_\tau| }(130)$&
82 &{9.5}&{5.0}&{4.0}&{3.7}&{3.5}\\
$\delta_{|d_\tau| }(30)$ & {11} &{4.5}&{3.3}&{3.0}&{2.9}&{2.9} \\
				\hline	
			\hline
		\end{tabular}
	\end{table}

	A color map of the precision that \(\mathrm{Re}(d_\tau)\) can be reached is plotted in Fig.~\ref{fig:three}. 
	Some 
	selected values are collected in Table~\ref{reedm}. 
	From the table, it is clear that the precision of \(\mathrm{Re}(d_\tau)\) is greatly improved at low energy when the spatial resolution $D$ becomes smaller. However, at high energy, the improvement is less significant, where \(\tau^{\pm}\) carry sufficient energy and can already fly far enough to be detected. 
	
	To study the precision of \(|d_\tau|\), we define 
	$\delta_{|d_\tau|}(D)^2
	=
	\delta_{\text{Im}}^2 
	+ \delta_{\text{Re}}(D)^2 
	.$
	In the last two rows of the table, we present the numerical values for the sensitivities for $|d_\tau|$ for $D= 130\mu m$ and $30\mu$m. Its precision
 at 6.3~GeV
	 can reach \(3.7 \times 10^{-18} e\text{cm}\) and \(2.9 \times 10^{-18} e\text{cm}\), respectively,  which is twice as good as the current experimental value reported in Eq.~\eqref{bel}.
	We note that after the upgrade of Belle II, the luminosity is expected to increase by two orders of magnitude~\cite{Belle-II:2010dht}, and therefore the precision could be  improved compared to the previous study~\cite{Belle:2021ybo}. However, it is important to highlight that in Ref.~\cite{Belle:2021ybo},  
	$
	\text{sgn} \left( (\hatl_- \times \hatl_+) \cdot \hatk \right)
	$
	was treated as a random number, and thus the results cannot be considered reliable.  
	The same selection method for $\tau$ described here can also be applied at Belle to resolve this issue.

	\section{Conclusion}
	
	Measuring \(d_\tau\) at future colliders offers a powerful probe of CP violation and potential new physics. We emphasize that reconstructing the full momentum of the \(\tau^-\) for \(\mathrm{Re}(d_\tau)\) is critical—an important aspect that has been previously overlooked. At the STCF, the precision for both 
   \(\mathrm{Re}(d_\tau)\) and 
	\(\mathrm{Im}(d_\tau)\) reaches its peak at a center-of-mass energy of \(6.3\,\mathrm{GeV}\), with attainable sensitivities of  
{ 	$
2.8  \times 10^{-18}\,e\,\mathrm{cm}$   and $0.7 \times 10^{-18}\,e\,\mathrm{cm},
	$ } 
	respectively, improving the current precision by approximately a factor of two. Near the \(\psi(2S)\) resonance, the achievable sensitivities are  
{ 		$
	1.8  \times 10^{-18}\,e\,\mathrm{cm}$}  and  $  11 \times 10^{-18}\,e\,\mathrm{cm}$
	for \(\mathrm{Im}(d_\tau)\) and \(\mathrm{Re}(d_\tau)\), respectively, indicating a moderate reduction in sensitivity at this energy.  
	On the other hand, BESIII has the capability to measure \(\mathrm{Im}(d_\tau)\) with a precision that may reach { \(1.0 \times 10^{-16}\,e\,\mathrm{cm}\)}.  
	These findings underscore the importance of careful momentum reconstruction and optimal energy selection for future \(\tau\)-EDM measurements.

	\section*{Acknowledgment}
	The authors would like to thank Hai-Bo Li and Xiaorong Zhou for discussions.
	This work was partially supported by the Fundamental Research Funds for the Central Universities, by NSFC grant numbers  
	12075299, 12090064, 
	12205063,  12375088 and W2441004, by National Key R\&D Program of China No. 2024YFE0109800. 
	
\end{document}